# Transition to turbulence through decline of viscosity

K.Y. Volokh[1]

*CEE, Technion - Israel Institute of Technology, Haifa 32000, Israel*

**Abstract**

Experiments (Mullin and Kreswell, 2005) show that transition to turbulence can start at Reynolds numbers lower than it is predicted by the linear stability analysis – the subcritical transition to turbulence. To explain these observations qualitatively we suggest that the onset of subcritical instability is related to decline of viscosity of the fluid: friction between fluid layers fails with the increase of the velocity gradient. Simply speaking, the faster a fluid layer travels with respect to the adjacent layer the lower is the friction between them. To describe the decline of friction theoretically we relax the assumption of the stability of the fluid material and introduce a constant of the *fluid strength*. Particularly, we enhance the Navier-Stokes model with a failure description by introducing the fluid strength in the constitutive equation for the viscous stress. The classical model is obtained from the enhanced one when the strength goes to infinity. We use the modified Navier-Stokes model to analyze the Couette flow between two parallel plates and find that the lateral perturbations can destabilize the flow and the critical Reynolds number is proportional to the fluid strength. The latter means that the classical Navier-Stokes model of a stable material with the infinite strength does not capture the subcritical transition to turbulence while the modified model does.

## 1. Introduction

Experiments (Mullin and Kreswell, 2005; Barkley and Tuckerman, 2005; Prigent and Dauchot, 2005) show that transition to turbulence can start at Reynolds numbers lower than it is predicted by the linear stability analysis (Schmid and Henningson, 2001) – the subcritical transition to turbulence. To explain these observations qualitatively we suggest that the onset of subcritical instability is related to decline of viscosity of the fluid: friction between fluid layers fails with the increase of the velocity gradient. Simply speaking, the faster a fluid layer travels with respect to the adjacent layer the lower is the friction between them. To describe the decline of friction theoretically we relax the assumption of the stability of the fluid material and introduce a constant of the *fluid strength*. Particularly, we enhance the Navier-Stokes model (Batchelor, 2000; Landau and Lifshitz, 1987) with a failure description by introducing the fluid strength in the constitutive equation for the viscous stress. The classical model is obtained from the enhanced one when the strength goes to infinity.

---
[1] E-mail: cvolokh@technion.ac.il

We use the modified Navier-Stokes model to analyze the Couette flow between two parallel plates and find that the lateral perturbations can destabilize the flow and the critical Reynolds number is proportional to the fluid strength. The latter means that the classical Navier-Stokes model of a stable material with the infinite strength does not capture the subcritical transition to turbulence while the modified model does.

**2. Constitutive law for viscous stress**

Momentum balance has the following form in the absence of body forces

$$\rho \frac{\partial \mathbf{v}}{\partial t} + \rho (\mathbf{v} \cdot \nabla) \mathbf{v} = \text{div}\, \boldsymbol{\sigma}, \qquad (1)$$

where $\rho$ is a constant mass density; $\mathbf{v}$ is a particle velocity; $t$ is time; and $\boldsymbol{\sigma}$ is the stress tensor.

In the case of incompressible fluid the stress can be specified as follows

$$\boldsymbol{\sigma} = -p\mathbf{1} + \boldsymbol{\tau}, \qquad (2)$$

$$\text{div}\, \mathbf{v} = 0, \qquad (3)$$

where $p$ is the Lagrange multiplier enforcing the incompressibility condition (3); $\mathbf{1}$ is the second-order identity tensor; and $\boldsymbol{\tau}$ is the so-called viscous stress.

Traditionally, the constitutive model for the viscous stress in Newtonian fluids is set as follows (Landau and Lifshitz, 1987; Batchelor, 2000)

$$\boldsymbol{\tau} = 2\eta \mathbf{D}, \tag{4}$$

where $\eta > 0$ is a viscosity coefficient and

$$\mathbf{D} = \frac{1}{2}(\nabla \mathbf{v} + \nabla \mathbf{v}^T) \tag{5}$$

is a symmetric part of the velocity gradient.

Substituting (2), (4), (5) in (1) we have the Navier-Stokes equations, which should be completed with boundary/initial conditions for velocities in order to set an initial-boundary-value problem (IBVP).

Instead of (4), however, we will use the following constitutive model enforcing failure of viscous/frictional bonds

$$\boldsymbol{\tau} = 2\eta \mathbf{D} \exp(-\varepsilon^2 / \phi^2), \tag{6}$$

$$\varepsilon^2 = \mathbf{D} : \mathbf{D} \equiv \mathrm{tr}(\mathbf{D}\mathbf{D}^T), \tag{7}$$

where $\varepsilon \geq 0$ is a scalar measure of the velocity gradient – the 'equivalent' velocity gradient and $\phi > 0$ is a constant of *fluid strength*, i.e. the maximum velocity gradient that preserves friction between the adjacent fluid layers.

There are two main modes for (6)

$$\boldsymbol{\tau} = 2\eta \mathbf{D} \quad \text{when} \quad \varepsilon \ll \phi, \tag{8}$$

$$\boldsymbol{\tau} = \mathbf{0} \quad \text{when} \quad \varepsilon \gg \phi. \tag{9}$$

The first mode (8) corresponds to the classical Navier-Stokes viscosity with the full internal friction while the second mode (9) corresponds to the loss of viscosity or internal friction. These two modes reflect upon Landau's notion that turbulent flows should be viscosity-free (Landau and Lifshitz, 1987)

In order to examine the linear stability of the flow we superimpose small motions on the existing ones and designate the perturbations with tildes. Varying equations (1)-(3) and (6) we have accordingly

$$\rho \frac{\partial \tilde{\mathbf{v}}}{\partial t} + \rho(\mathbf{v} \cdot \nabla)\tilde{\mathbf{v}} + \rho(\tilde{\mathbf{v}} \cdot \nabla)\mathbf{v} = \mathrm{div}\, \tilde{\boldsymbol{\sigma}}, \tag{10}$$

$$\tilde{\boldsymbol{\sigma}} = -\tilde{p}\mathbf{1} + \tilde{\boldsymbol{\tau}}, \quad \mathrm{div}\, \tilde{\mathbf{v}} = 0, \tag{11}$$

$$\tilde{\boldsymbol{\tau}} = 2\eta(\tilde{\mathbf{D}} - 2\mathbf{D}(\mathbf{D}:\tilde{\mathbf{D}})/\phi^2)\exp(-\mathbf{D}:\mathbf{D}/\phi^2), \tag{12}$$

The addition of the initial/boundary conditions of zero velocity perturbations completes the linearized IBVP.

Finalizing the general theoretical setting we notice that the classical Navier-Stokes theory is obtained from the above equations when the fluid strength goes to infinity: $\phi \to \infty$.

## 3. Plane Couette flow

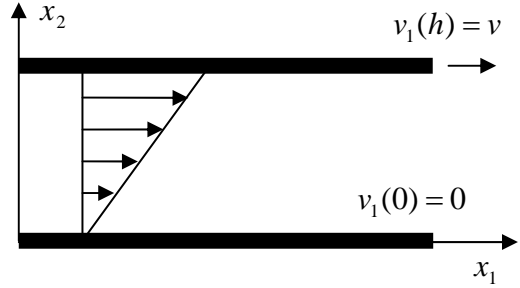

Fig.1 Flow between parallel plates.

Let us consider stability of the Plane Couettte flow – see also Drazin (2002) and Schmid and Nenningson (2001) for review. We assume that there is no pressure gradient and the velocity field has form: $\mathbf{v} = v_1(x_2)\mathbf{e}_1$, where $\mathbf{e}_1$ is a unit base vector (Fig.1). In this case we have

$$\mathbf{D} = D_{12}(\mathbf{e}_1 \otimes \mathbf{e}_2 + \mathbf{e}_2 \otimes \mathbf{e}_1), \quad D_{12} = \partial v_1 / \partial x_2, \tag{13}$$

$$\varepsilon = 2(\partial v_1 / \partial x_2)^2, \tag{14}$$

$$\boldsymbol{\tau} = \tau_{12}(\mathbf{e}_1 \otimes \mathbf{e}_2 + \mathbf{e}_2 \otimes \mathbf{e}_1), \quad \tau_{12} = 2\eta(\partial v_1 / \partial x_2)\exp(-2(\partial v_1 / \partial x_2)^2 / \phi^2), \tag{15}$$

and the reduced momentum balance (1) takes form

$$\frac{\partial \tau_{12}}{\partial x_2} = 0. \tag{16}$$

Substituting (15) in (16) and adding boundary conditions $v_1(0) = 0$ and $v_1(h) = v$ we find the following solution for velocity and stress fields

$$v_1 = vx_2 / h, \tag{17}$$

$$\sigma_{11} = \sigma_{22} = \sigma_{33} = -p; \quad \sigma_{12} = \sigma_{21} = (\eta v / h)\exp(-2(v/h)^2 / \phi^2). \tag{18}$$

Let us study the linear stability of the obtained solution. We assume that $\tilde{p} = 0$ and $\tilde{\mathbf{v}} = \tilde{v}_1(x_2)\mathbf{e}_1$. Then we have

$$\tilde{\mathbf{D}} = \frac{\partial \tilde{v}_1}{\partial x_2}(\mathbf{e}_1 \otimes \mathbf{e}_2 + \mathbf{e}_2 \otimes \mathbf{e}_1), \tag{19}$$

$$\tilde{\boldsymbol{\sigma}} = \tilde{\boldsymbol{\tau}} = \beta \frac{\partial \tilde{v}_1}{\partial x_2}(\mathbf{e}_1 \otimes \mathbf{e}_2 + \mathbf{e}_2 \otimes \mathbf{e}_1), \tag{20}$$

$$\beta = 2\eta(1 - \frac{4v^2}{\phi^2 h^2})\exp(-\frac{2v^2}{\phi^2 h^2}). \tag{21}$$

The momentum balance (10) reduces to

$$\rho \frac{\partial \tilde{v}_1}{\partial t} = \beta \frac{\partial^2 \tilde{v}_1}{\partial x_2^2}. \tag{22}$$

We further assume the following modes of the perturbed motion

$$\tilde{v}_1(x_2, t) = \text{constant} \cdot e^{\omega t} \sin(2\pi n x_2 / h), \quad n = 1, 2..., \tag{23}$$

where boundary conditions are obeyed: $\tilde{v}_1(x_2 = 1, h) = 0$; and $\omega$ is a real constant.

Substituting (21) and (23) in (22) we find

$$\omega = -\frac{8\eta\pi^2 n^2}{\rho h^2}(1 - \frac{4v^2}{\phi^2 h^2})\exp(-\frac{2v^2}{\phi^2 h^2}). \tag{24}$$

The Couette flow is stable when $\omega$ is negative and it loses stability when $\omega = 0$. The latter condition gives the critical velocity

$$v_c = \frac{h\phi}{2}, \tag{25}$$

and the critical Reynolds number

$$R_c = \frac{h\rho v_c}{\eta} = \frac{\rho h^2 \phi}{2\eta}. \tag{26}$$

It is interesting that in the case of the classical Navier-Stokes model where the fluid strength is infinite, $\phi \to \infty$, the flow is always stable with respect to the lateral perturbations – see also Romanov (1973), while in the case where strength is finite the flow can lose stability initiating the transition to turbulence.

**4. Conclusion**

In the present work we proposed a new explanation of the subcritical transition to turbulence. We assumed that the flow instability was triggered by the material instability of the fluid – the loss of internal friction. To dercribe it theoretically we induced a new constant of the fluid strength in the constitutive description of the fluid. The strength means the maximum 'equivalent' velocity gradient responsible for the viscosity/friction decline. We used the Navier-Stokes fluid model enhanced with the failure description for analysis of the plane Couette flow between two parallel plates. We found that the flow could lose its stability under the lateral velocity perturbations in the case where the fluid strength was finite. Our theoretical conclusions are in a good qualitative agreement with the experimental observations on the onset of the subcritical transition to turbulence reviewed recently in Mullin and Kerswell (2005). It is especially interesting that the localized turbulence patterns were observed in experiments, which could be theoretically interpreted as a local loss of internal friction due to the fluctuations in the fluid strength.